\documentclass[a4paper,fleqn,usenatbib]{mnras}
 \usepackage[normalem]{ulem}
\usepackage[T1]{fontenc}
\usepackage{ae,aecompl}
\usepackage{graphicx}	% Including figure files
\usepackage{amsmath}	% Advanced maths commands
\usepackage{amssymb}	% Extra maths symbols
\usepackage{multirow}
\usepackage{longtable}
\usepackage{natbib}
\hypersetup{draft}
\usepackage{rotating}
\title[Deuteration in cold cores]{Chemistry of TMC-1 with multiply deuterated species and spin chemistry of H$_2$, H$_2$$^+$, H$_3$$^+$ and their isotopologues} 

\author[Majumdar et al.]{
L. Majumdar$^{1,2}$\thanks{E-mail: liton.icsp@gmail.com}, P. Gratier$^{1}$, M. Ruaud$^{1}$, V. Wakelam$^{1}$, C. Vastel$^{3,4}$,   
\newauthor\ O. Sipil\"a$^{5}$, F. Hersant$^{1}$, A. Dutrey$^{1}$ , S. Guilloteau$^{1}$ %J.-C. Loison$^{4,5}$,  K. M. Hickson$^{4,5}$
\\ 
$^{1}$Laboratoire d'astrophysique de Bordeaux, Univ. Bordeaux, CNRS, B18N, allée Geoffroy Saint-Hilaire, 33615 Pessac, France\\
$^{2}$ Indian Centre For Space Physics, 43 Chalantika, Garia Station Road, Kolkata, 700084, India\\
$^{3}$Universit\'e de Toulouse, UPS-OMP, IRAP, Toulouse, France\\
$^{4}$CNRS, IRAP, 9 Av. Colonel Roche, BP 44346, F-31028 Toulouse Cedex 4, France\\
$^{5}$ Max-Planck-Institute for Extraterrestrial Physics (MPE), Giessenbachstr. 1, D-85748 Garching, Germany
}

\date{Accepted XXX. Received YYY; in original form ZZZ}

\pubyear{2015}

\begin{document}
%\label{firstpage}
\maketitle

\begin{abstract}

Deuterated species are unique and powerful tools in astronomy since they can probe the physical conditions, chemistry, and ionization level of various astrophysical media. Recent observations 
of several deuterated species along with some of their spin isomeric forms have rekindled the interest for more accurate studies on deuterium fractionation. This paper presents the first 
publicly available chemical network of multiply deuterated species along with spin chemistry implemented on the latest state-of-the-art gas-grain chemical code `NAUTILUS'.  
D/H ratios for all deuterated species observed at different positions of TMC-1 are compared with the results of our model, which considers multiply deuterated species along with the spin chemistry of light hydrogen bearing species 
H$_2$, H$_2$$^+$, H$_3$$^+$ and their isotopologues. We also show the differences in the modeled abundances of non-deuterated species after the inclusion of deuteration and spin chemistry in the model. Finally, we present a 
list of potentially observable deuterated species in TMC-1 awaiting detection.

\end{abstract}


%%%%%%%%%%%%%%%%%%%%%%%%%%%%%%%%%%%%%%%%%%%%%%%%%%

%%%%%%%%%%%%%%%%% BODY OF PAPER %%%%%%%%%%%%%%%%%%

\section{Introduction}
Until now (as of February 2016), almost 200 molecules have 
been detected in the interstellar medium or circumstellar shells (as listed by CDMS\footnote{\url{https://www.astro.uni-koeln.de/cdms/molecules}}). In addition, several deuterated species have been detected in various astrophysical media. Among them there were several 
detections:
\begin{itemize}
\item [-] in dark clouds: DCO$^+$ \citep{2009A&A...507..347V, 1977ApJ...217L.165G}, DNC~\citep{2009A&A...507..347V, 1978ApJ...225L..75T}, HDCO \citep{1985ApJ...299..947L}, 
D$_{2}$CO \citep{1990ApJ...362L..29T}; 
\item [-] in pre-stellar cores: D$_2$CO \citep{2004BaltA..13..402B}, 
H$_2$D$^+$~\citep{1999ApJ...521L..67S, 2003A&A...403L..37C, 2006ApJ...645.1198V, 2008A&A...492..703C, 2011A&A...526A..31P}, 
D$_2$H$^+$~\citep{2004ApJ...606L.127V, 2011A&A...526A..31P}, N$_2$D$^+$~\citep{2012A&A...538A.137M}, NHD$_2$~\citep{2000A&A...354L..63R}; 
\item [-] towards low-mass and high-mass protostars: D$_2$CO, HDCO~\citep{2011A&A...527A..39B}, DCOOCH$_3$~\citep{2010A&A...517A..17D, 2010ApJ...714.1120M}, 
HDO \citep{2012A&A...539A.132C, 2014MNRAS.445.1299C}, D$_2$O \citep{2010A&A...521L..31V}, OD \citep{2012A&A...542L...5P}; 
\item [-] in protoplanetary disks: DCN, 
DCO$^+$ \citep{2003A&A...400L...1V, 2006A&A...448L...5G, 2008ApJ...681.1396Q, 2015ApJ...810..112O}, 
N$_2$D$^+$ \citep{2015ApJ...809L..26H} and HD \citep{2013Natur.493..644B}, the main deuterium reservoir. 
\end{itemize}
The detection of ND$_3$ in the Barnard 1 cloud \citep{2002ApJ...571L..55L} and CD$_3$OH in IRAS 
16293-2422 \citep{2004A&A...416..159P} has 
shown the possibility of detecting multiply deuterated species in cold and high density regions. These observations opened new questions regarding the highest deuterium fractionation level that can occur in 
cold and high density regions. We have already entered into a new era of astrochemistry where powerful, high-sensitive, and high-resolution observational facilities like Atacama Large Millimetre Array (ALMA) are in operation. Thus, 
there is no doubt that many more singly and multiply deuterated species are going to be observed. By studying the chemistry of these deuterated species, we will be able to diagnose various cold and dense environments where stars are born (see for instance \citet{2013ApJS..207...27A} and references therein). 

Deuterium chemistry in the gas phase is believed to be controlled by the isotopic exchange reactions between HD and H$_3$$^+$, as the reservoir of deuterium is initially locked in the form of HD, via the reaction 
\begin{equation} 
\textrm{H}_3^+ + \textrm{HD} \rightarrow
\textrm{H}_2\textrm{D}^+ + \textrm{H}_2 + 232~\textrm{K}. 
\end{equation} 
 The above backward reaction is endothermic by 232 K and is thus negligible at low temperatures 
(less than 20 K). H$_2$D$^+$ can react with other abundant molecules (such as CO, N$_2$) to transfer D-atoms to other species \citep{2014prpl.conf..859C}. However, this reaction scheme becomes 
complicated if we consider nuclear spin states for the protonated species. Molecular hydrogen (H$_2$) has two distinct spin states under the permutation of identical protons. An ortho state (ortho-H$_2$) when the nuclear spin 
wave function is symmetric and a para state (para-H$_2$) when the nuclear spin wave function is antisymmetric.  The energy difference between the rotational ground states of ortho-H$_2$ and para-H$_2$ is 170.5 K \citep{2009JChPh.130p4302H}. 
%For example, molecular hydrogen (H$_2$) can exist in two distinct spin states where the nuclear spin wave 
%function is either symmetric (ortho-H$_2$) or antisymmetric (para-H$_2$). The difference in internal energy between the rotational ground states of ortho-H$_2$ and para-H$_2$ is 170.5 K \citep{2009JChPh.130p4302H}. 
Consequently if we consider the reaction (1) in backward direction with ortho-H$_2$ and ortho-H$_2$D$^+$, the reaction becomes 
exothermic by about 24 K with para-H$_3$$^+$ as one of the products. 
As a result, the ortho-to-para ratio of H$_2$ (opr(H$_2$) from now on in the text) becomes an important parameter in controlling the deuterium fractionation even at low temperature \citep{2006A&A...449..621F}.

The study of deuterium chemistry is a challenging task. The reason is the unavailability of any public database which provides a network with detailed deuterium chemistry along with 
 spin chemistry of important protonated species. In the past, many authors have studied deuterium fractionation  by cloning reactions involving hydrogen-bearing species 
 (\citet{2012ApJ...760...40A}, \citet{2013ApJS..207...27A}, \citet{2014ApJ...784...39A}, \citet{2015ApJ...808...21D}, \citet{2014A&A...562A..56M}, \citet{2014ApJ...782...73M}, \citet{2000A&A...361..388R}, 
 \citet{2000A&A...364..780R}, \citet{2013A&A...554A..92S}, \citet{2014ApJ...791....1T}). Among these studies,  \citet{2000A&A...361..388R,2000A&A...364..780R} reported one of the first gas-grain chemical models for deuteration, where the chemistry was limited to singly and a few doubly deuterated species along with surface chemistry only for H$_2$ and HD molecules. \citet{2012ApJ...760...40A} considered the chemistry of multiply deuterated species both in the 
 gas phase and on the grain surface but without considering spin-state chemistry. Similar models were also published in \citet{2013ApJS..207...27A}, \citet{2015ApJ...808...21D}, \citet{2014A&A...562A..56M, 2014ApJ...782...73M} where 
 multiple deuteration was taken into account without considering spin-state chemistry. \citet{2013A&A...554A..92S} and \citet{2014ApJ...791....1T} reported deuterium fractionation models with gas-grain chemistry of deuterated 
 species up to four atoms and the spin chemistry of H$_2$, H$_2$$^+$, H$_3$$^+$ and their deuterated isotopologues. More recently, \citet{2014ApJ...784...39A} have considered deuterated species with 7 atoms whereas \citet{2015A&A...578A..55S} 
 have considered 6 atoms. Both works include spin chemistry. In many cases, deuteration was studied by restricting the number of atoms in the deuterated species, without studying spin chemistry or grain chemistry in detail. The main goal of the present 
 paper is to present a more comprehensive deuterium fractionation model, which includes spin chemistry of H$_2$, H$_2$$^+$, H$_3$$^+$ and their deuterated isotopologues, the chemistry of multiply deuterated species and surface chemistry described in detail in Section 2. This network is the first of its kind to be made public ({\url{http://kida.obs.u-bordeaux1.fr/}}).
This paper is structured as follows. In Section 2, we present our new deuterium fractionation model. We give a detailed description of our deuterated network (deuspin.kida.uva.2016 from now in the rest of the text) along 
with its implementation in the gas-grain code NAUTILUS. We benchmark our model with the other published works. In Section 3, we discuss the general trends of deuterium chemistry. We compare our results with observations 
and discuss the effect of inclusion of deuspin.kida.uva.2016 on the predicted modelled abundances for several well observed non-deuterated species, and discuss some new observable deuterated species in dark clouds. Finally, in Section 4 
we draw our conclusions.

\section{Chemical model and network}

\subsection{The NAUTILUS chemical model}
To study deuteration together with spin chemistry, we use the NAUTILUS gas-grain chemical model \citep{2014MNRAS.440.3557R, 2015MNRAS.447.4004R, 2015MNRAS.453L..48W} under typical cold 
dense cloud conditions. NAUTILUS is a state-of-the-art chemical code, which can be applied to simulate various types of astronomical environments. Applications of this gas-grain chemical code have already been reported 
for dense clouds \citep{2015MNRAS.453L..48W}, low-mass protostellar envelopes \citep{2014MNRAS.441.1964B, 2016MNRAS.458.1859M}, and the outer regions 
of protoplanetary discs \citep{2011A&A...535A.104D}. NAUTILUS computes the abundances of species (e.g. atoms, ions, radicals, molecules) as a function of time in the gas phase and on the surfaces of interstellar grains. 
All the equations and the chemical processes included in the model are described in detail in \citet{2015MNRAS.447.4004R}. To include the effect of spin 
chemistry, we then modified all the variables in NAUTILUS relative to H$_2$ in terms of ortho-H$_2$ and para-H$_2$ (from now we will use o, p and m to represent ortho, para, and meta states 
\footnote{Here ortho, para and meta stand for different nuclear spin symmetries for a species. For example, H$_2$ has two spin states i.e. ortho (I=1 and g$_I$=3), para (I=0 and g$_I$=1) whereas D$_3$$^+$ has 
three spin states i.e. ortho (I=1,2 and g$_I$=16), para (I=0, g$_I$=1) and meta (I=1,3 and g$_I$=10). Among others, H$_2$$^+$, H$_3$$^+$, D$_2$$^+$, H$_2$D$^+$, D$_2$H$^+$ have only 
ortho and para spin states. Here, I and g$_I$ stand for nuclear spin angular momenta and nuclear spin statistical weights respectively. See  \citet{2009JChPh.130p4302H} for more details.} in the rest of the text). 

In the model,  several types of chemical reactions are considered in the gas phase by following the kida.uva.2014 chemical network of \citet{2015ApJS..217...20W}. These reactions can be classified in four categories: 
bimolecular reactions between neutral species, between charged species and between neutral and charged species, unimolecular reactions (i.e. photoreactions with direct UV photons and UV photons produced by the 
deexcitation of H$_2$ excited by the cosmic ray particles), and direct ionisation and dissociation by cosmic ray particles. Here the interstellar ice is modelled by a one phase rate equation approach \citep{1992ApJS...82..167H}, i.e. 
there is no differentiation between the species in the mantle and on the surface. In our model, the gas and the grains are coupled to each other via four interaction processes : physisorption of gas phase species onto grain surfaces, 
diffusion of the accreted species, reaction at the grain surface, and evaporation to the gas phase. These evaporation processes can also be of various types. The evaporation mechanisms considered here are: thermal 
(which are inefficient at dense cloud conditions), induced by cosmic-rays \citep {1993MNRAS.261...83H}, and chemical as defined by \citet{2007A&A...467.1103G}. According to \citet{2014MNRAS.445.2854W}, photodesorption 
induced by cosmic-ray secondary photons is less efficient than chemical desorption. This is why we did not consider it in our model.  Any species can diffuse by thermal hopping only with 
a barrier of 0.5$\times$E$_{\rm D}$ where E$_{\rm D}$ is the species binding energy. By following \citet[]{2007A&A...467.1103G}, we consider that approximately 1\% of the products is allowed to desorb due to 
chemical desorption. We do not take into account the cosmic-ray induced diffusion mechanism (CRD)  since for high visual extinction, effect of CRD is negligible  \citep[see][]{2014MNRAS.440.3557R} \\

%We have not introduced photodesorption 
%induced by cosmic-ray secondary photons since it was found to be inefficient compared to the chemical desorption mechanism by \citet{2014MNRAS.445.2854W}. Any species can diffuse by thermal hopping only with 
%a barrier of 0.5$\times$E$_{\rm D}$ (with E$_{\rm D}$ the species binding energy). For chemical desorption, i.e. partial desorption of the products due to the exothermicity of the reactions occurring at the surface of the grains, 
%approximately 1\% of the products is allowed to desorb \citep[see][]{2007A&A...467.1103G} . 
%The Cosmic-Ray Induced Diffusion mechanism is not included because it does not have any impact on the surface chemistry at high 
%visual extinction \citep[see][]{2014MNRAS.440.3557R}. \\

To simulate deuterium chemistry together with spin chemistry in dark clouds, the model is 
used with homogeneous physical conditions and integrated over $10^7$~yrs. The initial elemental abundances reported in Table 1 are the same as in \citet{2011A&A...530A..61H} with deuterium and fluorine elemental abundances
 relative to hydrogen of $1.6\times 10^{-5}$ \citep{2006ApJ...647.1106L} and $6.68\times 10^{-9}$ \citep{2005ApJ...628..260N} respectively. The species are assumed to be initially in an atomic form as in diffuse clouds except for hydrogen
and deuterium, which are initially in H$_2$ and HD forms respectively. All elements (e.g. C, S, Si, Fe, Na, Mg, Cl, and P) with an ionization potential lower than 13.6 eV are initially singly ionised. For our standard model, we have used a C/O ratio of 0.7 (i.e. the oxygen elemental abundance is $2.4\times 10^{-4}$). The ortho-to-para H$_2$ ratio is initially set to its statistical value of 3.  The model was run with a dust and gas temperature of 10~K, a total proton density of $2\times 10^4$~cm$^{-3}$, a cosmic-ray ionization rate of $1.3\times 10^{-17}$~s$^{-1}$, and a visual extinction of 30 mag. 

\begin{table*}
 \centering
 %\begin{minipage}{19cm}
  \caption{Initial abundances used in our model.}
  \begin{tabular}{ll}
  \hline \hline
   Element & Abundance relative to H\\
\hline
o-H$_{2}$& {3.75$\times10^{-1}$} \\
p-H$_{2}$& {1.25$\times10^{-1}$}  \\
He&{9$\times10^{-2}$}\\
  % \footnote{See discussion in \citet{Wakelam&Herbst2008}.}}\\
N&{6.2$\times10^{-5}$}\\
 %\footnote{\label{bb}\citet{Jenkins2009}.}}\\
O&{2.4$\times10^{-4}$}\\
 % \footnote{\label{cc}See discussion in \citet{Hincelin2011}.}}\hspace{1cm}(C/O=0.5)\\
%&{1.4$\times10^{-4}$
  %\footref{cc}}\hspace{1cm}(C/O=1.2)\\
C$^{+}$&{1.7$\times10^{-4}$}\\
  % \footref{bb}}\\
S$^{+}$&{8$\times10^{-8}$}\\
    %\footnote{\label{dd}Low metal abundances \citep{Graedel1982}.}}\\
Si$^{+}$&{8$\times10^{-9}$}\\
   %\footref{dd}}\\
Fe$^{+}$&{3$\times10^{-9}$}\\
   %\footref{dd}}\\
Na$^{+}$&{2$\times10^{-9}$}\\
   %\footref{dd}}\\
Mg$^{+}$&{7$\times10^{-9}$}\\
   %\footref{dd}}\\
P$^{+}$&{2$\times10^{-10}$}\\
   %\footref{dd}}\\
Cl$^{+}$&{1$\times10^{-9}$}\\
   %\footref{dd}}\\
 F&{6.68$\times10^{-9}$}\\
   %\footref{dd}}\\  
 HD&{1.60$\times10^{-5}$}\\ 
 \hline
%\label{table1}
\end{tabular}
%\end{minipage}
\end{table*}

\begin{figure}
%\centering
%\includegraphics[width=0.45\textwidth, angle=0]{./fig1.pdf}
\includegraphics[width=0.5\textwidth]{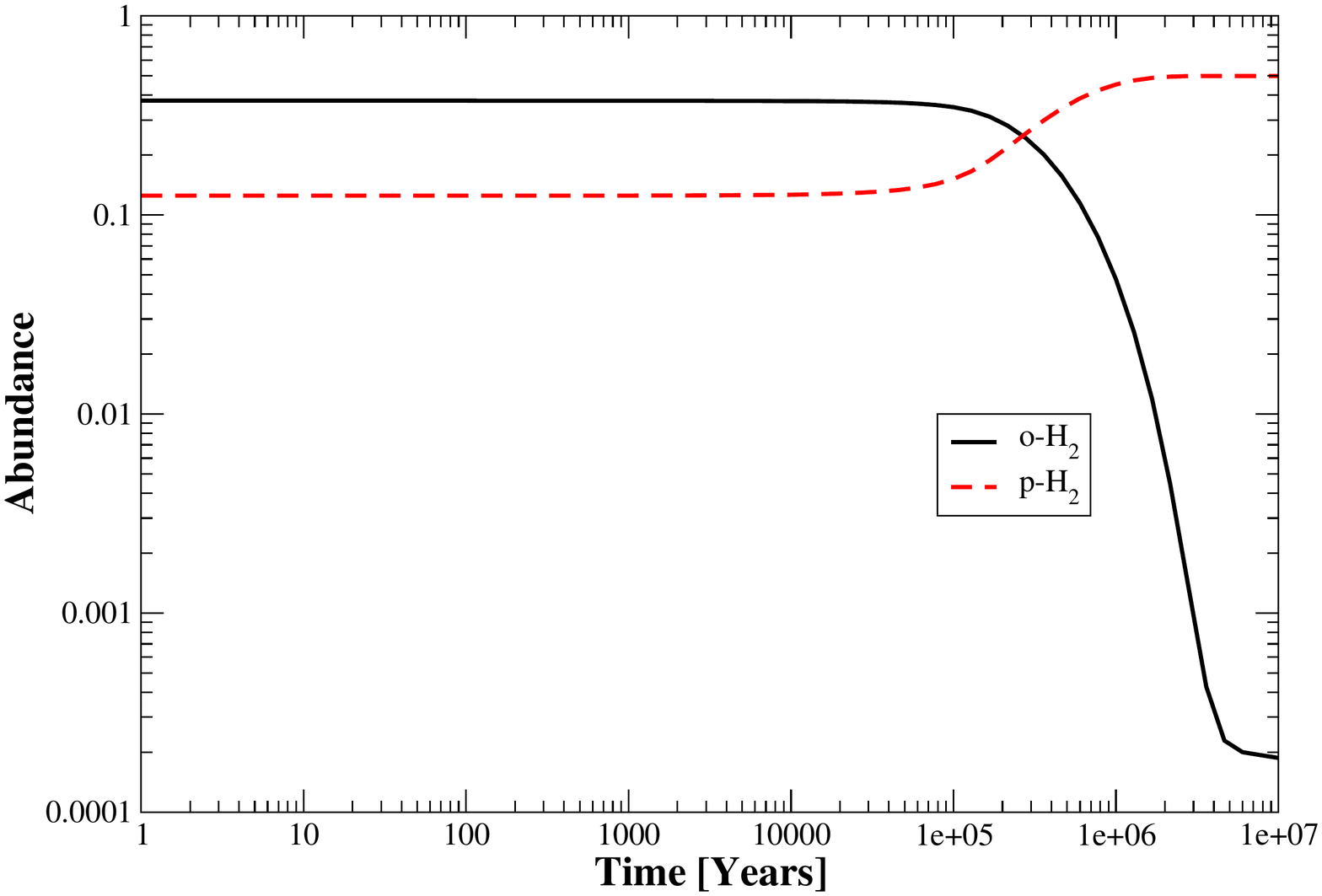}
\caption{ Gas-phase abundance of ortho and para H$_2$ with respect to n$_H$ as a function of time.}
\label{fig:diffDist}
\end{figure}

\begin{figure}
\centering
\includegraphics[width=0.5\textwidth]{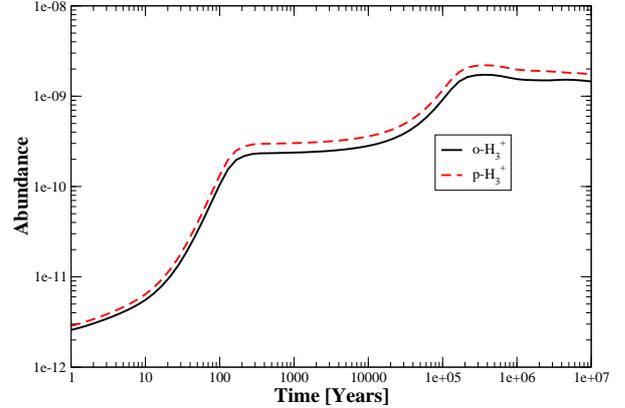}
\caption{Gas-phase abundance of ortho and para forms of H$_3$$^+$ with respect to n$_H$ as a function of time.}
\label{fig:diffDist}
\end{figure}

\begin{figure}
\centering
\includegraphics[width=0.5\textwidth]{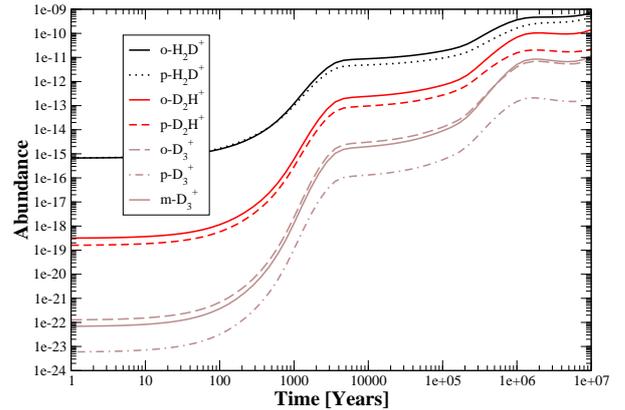}
\caption{Gas-phase abundance of various isotopologues of H$_3$$^+$ with respect to n$_H$ fas a function of time.}
\label{fig:diffDist}
\end{figure}

\subsection{Developed chemical network from KIDA 2014}
The initial gas-phase chemical network was adapted from the kida.uva.2014 network\footnote{\url{http://kida.obs.u-bordeaux1.fr/}} (\citet{2015ApJS..217...20W}, which includes 489 species 
composed of 13 elements (H, He, C, N, O, Si, S, Fe, Na, Mg, Cl, P, F) linked with 7509 reactions. Our starting network includes various updates of the HCN/HNC chemistry by \citet{2014MNRAS.443..398L}, carbon 
chemistry by \citet{2014MNRAS.437..930L}, branching ratios for reactions forming C$_{n=2-10}^{(0,+)}$, C$_{n = 2- 4}$H$^{(0, +)}$, and C$_3$H$_2^{(0,+)}$ from \citet{2013ApJ...771...90C} and also various new data 
sheets from KIDA database. This network has been extended to include the spin states of H$_2$, H$_2$$^+$ and H$_3$$^+$ and their deuterated isotopologues.   

To add the spin states of H$_2$, H$_2$$^+$, H$_3$$^+$  into kida.uva.2014, we applied the method described by \citet{2015A&A...578A..55S} in which the branching ratios are calculated using 
Oka's method \citep{2004JMoSp.228..635O}. In this method, selection rules for reactive collisions involving different spin species are derived using angular momentum algebra. Branching ratios resulting from this method 
correspond to pure nuclear spin statistical weights under the assumption that the nuclei are completely mixed in the reaction. Following \citet{2015A&A...578A..55S}, we applied the same method to all reactions except charge-transfer 
reactions, where we assume that spin states are conserved. We have added an activation energy of 170 K to the $\gamma$ coefficient (of the temperature dependent rate coefficients $\rm k(T) = \alpha (T/300)^{\beta} e^{-\gamma/T}$) for the reactions where o-H$_2$ is formed (see \citet{2014ApJ...784...39A} for similar methods). The formation of o-H$_2$ and p-H$_2$ in reactions that involve species, whose spin states are not tracked (i.e., other than H$_2$, H$_2$$^+$, and H$_3$$^+$), is handled with the ``recombination'' approach described in detail in Sect. 2.3.1 of \citet{2015A&A...578A..55S}. Our deuteration routine is similar to the one described in \citet{2013ApJS..207...27A} where deuterons are substituted for protons in the 
reactions, and branching ratios are calculated assuming complete scrambling. Here, we do not consider the spin chemistry of multiply deuterated species. For multiply deuterated species, the calculation of nuclear spin-state 
branching ratios is more complicated than for hydrogenated species since there is no one-to-one correspondence between angular momentum and symmetry representations. In the present work, we include the spin-state 
chemistry of the deuterated isotopologues involved in the H$_3$$^+$ + H$_2$ reaction system (which is essential in the present context) from Table III and IV of  \citet{2009JChPh.130p4302H} complemented by important reactions involving 
light hydrogen and deuterium-bearing species 
(H$_3$$^+$, H$_2$D$^+$, HD etc.) from \citet{2004A&A...427..887F} and \citet{2004A&A...418.1035W}. We also included dissociative recombination reactions for H$_3$$^+$ and its deuterated forms 
from \citet{2009A&A...494..623P}. We have added all these important reactions as a supplementary file in the format of kida.uva.2014 network. In addition, we have added several important reactions from the literature: important reactions for deuterations from \citet{2013ApJS..207...27A} and water deuteration chemistry from \citet{Talukdar1996177}, \citet{1999ApJ...510L.145B}. The entire network considered here is available on the KIDA\footnote{\url{http://kida.obs.u-bordeaux1.fr/}} website. 

Our network for surface reactions and gas-grain interactions is based on the one from \citet{2007A&A...467.1103G} with several additional processes from \citet{2015MNRAS.447.4004R}. We extended our network to include deuteration by assuming statistical branching ratios (see for instance \citet{2014ApJ...791....1T}  for the same method). 
According to \citet{2013A&A...550A.127T}, although deuterated isotopologues have the higher mass, they seem to show similar binding energies to their hydrogenated counterpart. Therefore, by following these studies and also 
due to the unavailability of binding energies for deuterated species, we assume the same binding energies for all deuterated and non-deuterated species.  Finally, we have nearly 7700 reactions on grain surface linked with the 111000 reactions 
in gas phase. If we sum up all these modifications, the deuspin.kida.uva.2016 network has the following characteristics: 

\begin{itemize}
\item [-] Spin chemistry of light hydrogen bearing species (H$_2$, H$_2$$^+$, H$_3$$^+$) essential to take into account the deuterium fractionation properly. 
\item[-] Spin chemistry of H$_3$$^+$ + H$_2$ reacting system and their isotopologues from  \citet{2009JChPh.130p4302H}. 
\item[-] Chemistry of multiply deuterated species.
\item[-] Extension of the kida.uva.2014 network to include full deuterium fractionation for species containing any of the 13 elements (H, He, C, N, O, Si, S, Fe, Na, Mg, Cl, P, F).
\item[-] All reactions are written in the the same format as in the KIDA database for the different types so that in future users can export this network easily without any confusion.
\end{itemize}

\subsection{Benchmarking spin and deuterium chemistry}
H$_2$, D$_2$, H$_3$$^+$, H$_2$D$^+$, D$_2$H$^+$ and D$_3$$^+$ along with their spin isomers are the main species that dictate deuterium fractionation at low temperature 
(see for instance \citet{2014prpl.conf..859C} and references therein). In addition, deuterium fractionation strongly depends on the initial  opr(H$_2$) considered. Ortho and para H$_2$ are formed on the surfaces of interstellar grains with a statistical ratio of 3:1 \citep{2010ApJ...714L.233W} and proton-exchange reactions in the gas phase then convert ortho-H$_2$ into para-H$_2$ (see \citet{2014prpl.conf..859C} and references therein for more discussions). In our model, we set an initial ratio of 3:1, in agreement with the experimental findings of \citet{2010ApJ...714L.233W}. 

Before applying our model to a dense core (e.g. a TMC-1 like environment), we have benchmarked our model with other published works. The importance of spin-state chemistry is now widely 
accepted and has been tested in different stages of star formation process: diffuse clouds \citep{2014ApJ...787...44A}, starless/prestellar cores 
(\citet{2009A&A...494..623P}, \citet{2013A&A...554A..92S}) and protostellar systems (\citet{2013A&A...550A.127T}). But in all cases spin-state chemistry was discussed to address particular issues by adopting various types of physical and chemical model making a direct comparison with our model very difficult. Two things can be compared though: the time scale for gas phase ortho-H$_2$ to para-H$_2$ conversion and abundance profiles for the two spin 
isomers of H$_3$$^+$. Fig. 1 shows the evolution of the ortho and para forms of H$_2$ in the gas-phase under dense core conditions. In our model, opr(H$_2$) becomes unity at around few times 10$^5$ years and drops down to 10$^{-3}$ or below after this. This is consistent 
with the result of \citet{2013A&A...554A..92S}. Fig. 2 shows the evolution of the ortho and para forms of H$_3$$^+$. At 10$^7$ year, the abundances of o-H$_3$$^+$ and p-H$_3$$^+$ are respectively $1.45\times10^{-09}$  and $1.73\times10^{-09}$. 
\citet{2015A&A...578A..55S} reported $2.00\times10^{-09}$ for o-H$_3$$^+$ under the same conditions.    

To validate deuterium chemistry with associated spin chemistry, one basic test would be to compare the abundance profile of  H$_2$D$^+$, D$_2$H$^+$ and D$_3$$^+$ with other models, since they are the major 
species controlling the deuterium fractionation. In order to perform a more realistic comparison, we ran our model with a similar 
initial opr(H$_2$)=10$^{-3}$ (i.e kinetically equilibrated H$_2$ at low temperature, see \citet{2013ApJ...770L...2F}) mentioned in \citet{2015A&A...578A..55S}.  Fig. 3 shows the abundances of all the isotopologues of H$_3$$^+$. At 10$^7$ year, the abundances of o-H$_2$D$^+$, p-H$_2$D$^+$, o-D$_2$H$^+$, 
p-D$_2$H$^+$, o-D$_3$$^+$, p-D$_3$$^+$, m-D$_3$$^+$ are respectively $7.0\times10^{-10}$,  $4.7\times10^{-10}$, $1.4\times10^{-10}$, $2.2\times10^{-11}$, $8.5\times10^{-12}$, $2.1\times10^{-13}$, $1.1\times10^{-11}$. In 
Fig. 3 of \citet{2015A&A...578A..55S}, at 10$^7$ year the reported abundances of o-H$_2$D$^+$ and p-D$_2$H$^+$ are respectively  $4.0\times10^{-10}$  and $1.0\times10^{-11}$ under the same conditions, within a 
factor of 2 in agreement with our model. These differences are quite minor considering the differences in the starting networks (our work is based on kida.uva.2014 network whereas \citet{2015A&A...578A..55S} used osu\_03\_2008 network). But we have found the similar trend for all the isotopologues of H$_3$$^+$ as described in \citet{2015A&A...578A..55S} i.e. abundances are much higher at late times and this is attributed 
due to the depletion of their main reaction partners (e.g., CO) onto grain surfaces.

%{\bf These minor differences in abundances are obvious since the starting networks and the models are different, but the result is globally similar.}
%\citep[see][]{2014MNRAS.440.3557R}
\begin{figure*}
\centering
\includegraphics[width=\textwidth]{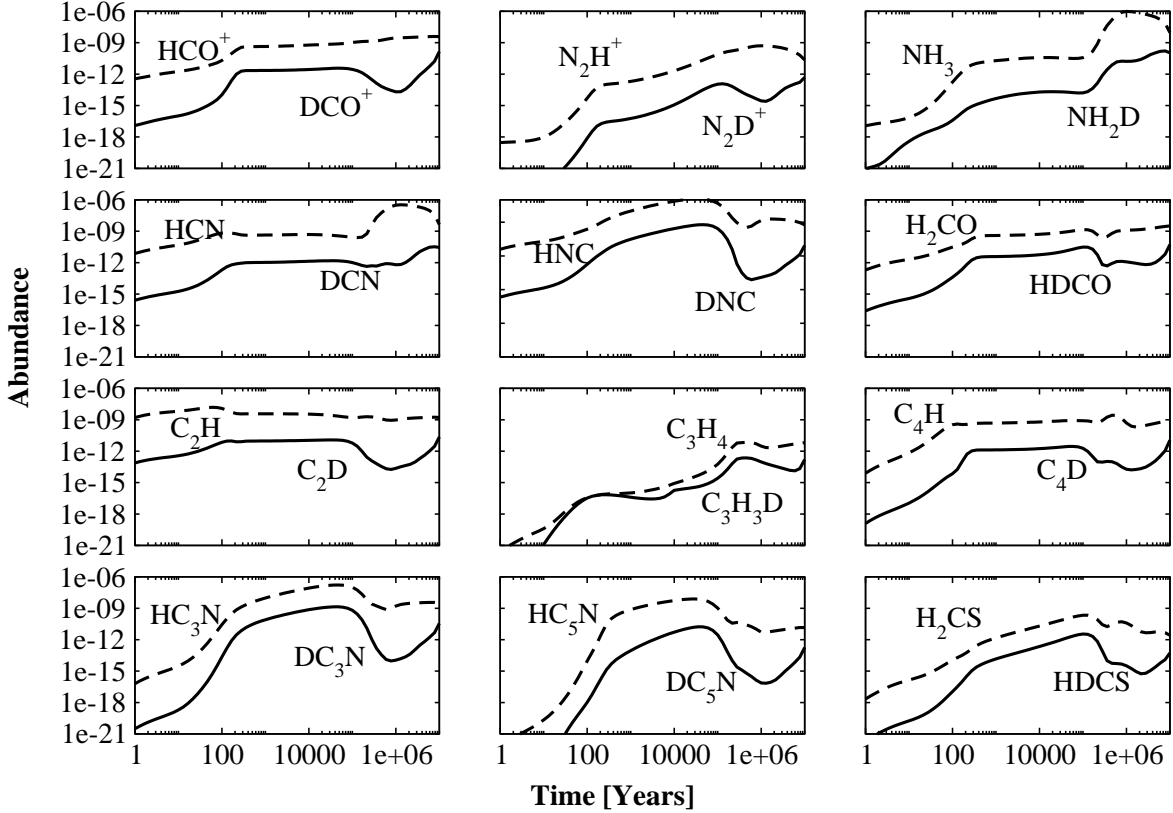}
\caption{Abundance with respect to n$_H$ as a function of time of a selection of deuterated species observed in dense cores. Dotted lines correspond to non-deuterated species and solid lines correspond to deuterated species.}
\label{fig:diffDist}
\end{figure*}

\section{Results and discussion}

\subsection{General trends of deuterium chemistry}
Using the model described above, we have investigated general trends of deuterium chemistry in the ISM. For this purpose, we selected two deuterated ions (DCO$^+$, N$_2$D$^+$) and 
several deuterated neutrals (NH$_2$D, DCN, DNC, HDCO, C$_2$D, C$_3$H$_3$D, C$_4$D, DC$_3$N, DC$_5$N, HDCS) as they have been observed in TMC-1 \citep[see for instance ][and references therein]{2013ApJS..207...27A}. The 
abundances predicted by our model are shown in Fig. 4 as a function of time. Our model shows that the abundance of HCO$^+$ increases with time and its formation is 
very efficient after 10$^2$ years due to the reaction between HOC$^+$ and o-H$_2$. Around this time, most of the gas phase H$_2$ is in the form of o-H$_2$. For DCO$^+$, we observe a very rapid fall in abundance between 10$^5$ and 10$^6$ years.  Around this time, H$_2$O becomes an efficient destruction partner of DCO$^+$ since the formation of H$_2$O is very efficient (due to its highly abundant precursors H$_3$O$^+$ and o-H$_2$). After 10$^6$ years, we observe an increase in abundance for DCO$^+$ due its efficient formation via the reaction o-H$_2$D$^+$ + CO. Since around this time o-H$_2$D$^+$
becomes highly abundant due to low opr(H$_2$). For the other deuterated ion N$_2$D$^+$, we 
observe a similar trend as with DCO$^+$. Between 10$^5$ and 10$^6$ years,  N$_2$D$^+$ is destroyed efficiently by H$_2$O to form H$_2$DO$^+$. Similarly, after 10$^6$ years, rapid formation of N$_2$D$^+$ is due to the 
efficient formation via the reaction of o-H$_2$D$^+$ + N$_2$.  

Among the deuterated neutrals, we observe a slightly different behaviour for NH$_2$D and C$_3$H$_3$D as compared to deuterated ions. For NH$_3$ and NH$_2$D, we find an increase in the abundances after 10$^5$ 
years. Around this time both NH$_3$ and NH$_2$D are formed via the dissociative recombination reactions of NH$_4$$^+$ and NH$_3$D$^+$. Both these ions are formed back again from various ion-molecular destruction reactions 
of NH$_3$ and NH$_2$D by the ions HCO$^+$, H$_3$O$^+$. After 10$^6$ years, the abundance of NH$_3$ drops quickly due its efficient depletion onto grain surface.    
But C$_3$H$_4$ and C$_3$H$_3$D show an increase 
in abundance in between 10$^5$ and 10$^6$ years due to their efficient formation via the barrier-less surface reactions s-H + s-C$_3$H$_3$ and s-H + s-C$_3$H$_2$D (here, `s' represents species on the surface of grains). 
 For the rest of the deuterated neutrals, we observe a general trend between 10$^5$ and 10$^6$ years i.e. abundances decrease. DNC, which is an isomer of DCN, shows a decrease in 
abundance around this time by several orders of magnitude. This is due to the efficient destruction of DNC by highly abundant p-H$_3$$^+$. HCN shows a peak in the abundance at 10$^6$ year due to its high abundant precursors H$_2$CN$^+$ and 
p-H$_2$. HDCO is less destroyed than the other species at later times because its main destruction path is O + HDCO. DC$_3$N, DC$_5$N, HDCS, on the 
contrary, show a rapid fall in the abundance in between 10$^5$ and 10$^6$ years due to their efficient destruction by highly abundant p-H$_3$$^+$ and HCO$^+$. In Table 2, we have listed major reactions of 
production and destruction for these species at high and low opr(H$_2$) regimes.

\begin{table*}
%\begin{sidewaystable}
\caption{Main reactions of production and destruction for observed deuterated species in TMC-1 cloud in the high and low opr(H$_2$) regimes. 
Here $10^5$~yr corresponds to high opr(H$_2$) and $10^6$~yr corresponds to low opr(H$_2$) regimes.}
%\scriptsize
\tiny
\hskip -1.0cm
%\begin{center}
\begin{tabular}{|l|c|c|c|c|}
\hline
\hline
%& \multicolumn{2}{|c|}{$10^5$~yr}& \multicolumn{2}{|c|}{$10^6$~yr} \\

Species & Formation & Destruction & Formation & Destruction \\
             &($10^5$~yr)&($10^5$~yr) & ($10^6$~yr)& ($10^6$~yr)\\  

\hline
NH$_2$D & NH$_3$D$^+$ + e$^{-}$ $\rightarrow$ H+ NH$_2$D & H$_3$O$^+$ + NH$_2$D $\rightarrow$  NH$_3$D$^+$ + H$_2$O   & NH$_3$D$^+$ + e$^{-}$ $\rightarrow$ H + NH$_2$D   &     

H$^+$ + NH$_2$D $\rightarrow$ NH$_2$D$^+$ + H  \\

 &  &  HCO$^+$ + NH$_2$D $\rightarrow$ NH$_3$D$^+$ + CO       &      & HCO$^+$ + NH$_2$D $\rightarrow$ NH$_3$D$^+$ + CO                      \\

\hline
HDCO & O + CH$_2$D $\rightarrow$ HDCO + H & O + HDCO $\rightarrow$ H + CO + OD & s-H + s-DCO$\rightarrow$ HDCO             &   HCO$^+$ + HDCO$\rightarrow$ H$_2$DCO$^+$ + CO       \\
&   & O + HDCO $\rightarrow$ D + CO + OH&    &    p-H$_3$$^+$ + HDCO$\rightarrow$ H$_3$CO$^+$ + HD    \\

\hline
DCN & CH$_2$D$^+$ + HC$_3$N $\rightarrow$ DCN + c-C$_3$H$_3$$^+$  & CH$_3$$^+$ + DCN $\rightarrow$ C$_2$H$_3$DN$^+$ &    HDCN$^+$ + p-H$_2$$\rightarrow$p-H$_3$$^+$ + DCN &   
 HCO$^+$ + DCN $\rightarrow$ HDCN$^+$ + CO   \\

& N + DCO $\rightarrow$ DCN + O &   &   &  p-H$_3$$^+$ + DCN $\rightarrow$ H$_2$CN$^+$ + HD        \\
& N + CHD $\rightarrow$ DCN + H & && o-H$_3$$^+$ + DCN $\rightarrow$ HDCN$^+$ + o-H$_2$ \\
 
 \hline
 DNC & N + CHD $\rightarrow$ DNC + H & p-H$_3$$^+$ + DNC $\rightarrow$ H$_2$CN$^+$ + HD &   C + NHD $\rightarrow$ DNC + H      &   HCO$^+$ + DNC $\rightarrow$ HDCN$^+$ + CO              \\
 & & HCO$^+$ + DNC $\rightarrow$ HDCN$^+$ + CO   & & p-H$_3$$^+$ + DNC $\rightarrow$ H$_2$CN$^+$ + HD\\
 &&  o-H$_3$$^+$ + DNC $\rightarrow$ HDCN$^+$ + o-H$_2$    & &  o-H$_3$$^+$ + DNC $\rightarrow$ HDCN$^+$ + o-H$_2$                \\

 \hline
 C$_2$D & C + CHD $\rightarrow$ C$_2$D + H & O + C$_2$D $\rightarrow$ CD + CO&   C$_2$H$_2$D$^+$ + e$^-$ $\rightarrow$ C$_2$D + H + H  &   C$_2$D + C$_3$$\rightarrow$ C$_5$ + D \\
 & C +C$_3$D$\rightarrow$ C$_2$D + CO & C$_2$D + C$_3$$\rightarrow$C$_5$ +D  &&\\

 \hline
 C$_4$D & C + c-C$_3$HD $\rightarrow$ C$_4$D + H & C + C$_4$D $\rightarrow$ C$_5$+ D &   C + c-C$_3$HD $\rightarrow$ C$_4$D + H       &  HCO$^+$ + C$_4$D$\rightarrow$ C$_4$HD$^+$ + CO \\
 && O + C$_4$D $\rightarrow$ C$_3$D+ CO &          &  p-H$_3$$^+$ + C$_4$D$\rightarrow$ C$_4$H$_2$$^+$ +HD                \\ 
 &&&& o-H$_3$$^+$ + C$_4$D $\rightarrow$ C$_4$H$_2$$^+$ +HD \\
  
 \hline
 DCO$^+$ & HCO$^+$ + D $\rightarrow$ DCO$^+$ + H & DCO$^+$ + H$_2$O $\rightarrow$ H$_2$DO$^+$ + CO &  o-H$_2$D$^+$ + CO$\rightarrow$ DCO$^+$ + o-H$_2$        &   DCO$^+$ + e$^-$ $\rightarrow$ CO + D \\
 & CH$_2$D$^+$ + O$\rightarrow$ DCO$^+$ + o-H$_2$ & DCO$^+$ + C $\rightarrow$ CD$^+$ + CO &   HCO$^+$ + D $\rightarrow$ DCO$^+$ + H & DCO$^+$ + H$_2$O $\rightarrow$ H$_2$DO$^+$ + CO   \\
 &&  DCO$^+$ + HCN $\rightarrow$ HDCN$^+$ + CO\\

 \hline
  N$_2$D$^+$ & N$_2$H$^+$ + D $\rightarrow$ N$_2$D$^+$ + H & N$_2$D$^+$ + H$_2$O $\rightarrow$ H$_2$DO$^+$ + N$_2$ &    o-H$_2$D$^+$ + N$_2$$\rightarrow$ N$_2$D$^+$ + o-H$_2$ & N$_2$D$^+$ + CO$
  \rightarrow$ DCO$^+$ + N$_2$               \\
   && N$_2$D$^+$ + CO $\rightarrow$ DCO$^+$ + N$_2$ & p-H$_2$D$^+$ + N$_2$$\rightarrow$ N$_2$D$^+$ + p-H$_2$&         \\
  &&&   N$_2$H$^+$ + D$\rightarrow$ N$_2$D$^+$ + H &   \\ 
  
 \hline
  c-C$_3$HD& c-C$_3$H$_2$D$^+$ + e$^-$ $\rightarrow$ c-C$_3$HD + H & C + c-C$_3$HD $\rightarrow$ C$_4$D + H &    H + C$_3$H$_3$D$\rightarrow$ c-C$_3$HD + o-H$_2$    & HCO$^+$ + c-C$_3$HD$\rightarrow$C$_3$H$_2$D$^+$ + CO                         \\
  && C + c-C$_3$HD $\rightarrow$ C$_4$H + D &H + C$_3$H$_3$D$\rightarrow$ c-C$_3$HD + p-H$_2$ &  p-H$_3$$^+$$\rightarrow$ C$_3$H$_3$$^+$ + HD \\  
  
 \hline
   C$_3$H$_3$D& s-H + s-C$_3$H$_2$D $\rightarrow$ C$_3$H$_3$D & C + C$_3$H$_3$D $\rightarrow$ C$_4$H$_2$D + H & s-H + s-C$_3$H$_2$D $\rightarrow$ C$_3$H$_3$D &    C + C$_3$H$_3$D $\rightarrow$ C$_4$H$_2$D + H                                     \\
   & & C + C$_3$H$_3$D $\rightarrow$ C$_4$H$_2$ + HD &      &     C + C$_3$H$_3$D $\rightarrow$ C$_4$H$_2$ + HD        \\   
   
 \hline
  DC$_3$N & C + CHDCN $\rightarrow$ DC$_3$N + H & p-H$_3$$^+$ + DC$_3$N $\rightarrow$ HC$_3$NH$^+$ + HD &       C + CHDCN $\rightarrow$ DC$_3$N + H   & HCO$^+$ + DC$_3$N$\rightarrow$ C$_3$HDN$^+$ + CO    \\
  && o-H$_3$$^+$ + DC$_3$N $\rightarrow$ HC$_3$NH$^+$ + HD &  C$_3$HDN$^+$ + e$^-$$\rightarrow$  DC$_3$N + H &    p-H$_3$$^+$ + DC$_3$N $\rightarrow$ HC$_3$NH$^+$ + HD                \\
  \hline
DC$_5$N & C$_5$H$_2$DN$^+$ + e$^-$ $\rightarrow$ DC$_5$N + o-H$_2$ & p-H$_3$$^+$ + DC$_5$N $\rightarrow$ H$_2$C$_5$N$^+$ + HD &     CN + C$_4$HD$\rightarrow$ DC$_5$N + H      & 
HCO$^+$ + DC$_5$N$\rightarrow$  C$_5$HDN$^+$ + CO           \\
&&  o-H$_3$$^+$ + DC$_5$N $\rightarrow$ H$_2$C$_5$N$^+$ + HD &     &     p-H$_3$$^+$ + DC$_5$N $\rightarrow$ H$_2$C$_5$N$^+$ + HD             \\
\hline
HDCS & S + CH$_2$D $\rightarrow$ H + HDCS & p-H$_3$$^+$ + HDCS $\rightarrow$ H$_3$CS$^+$ + HD & S + CH$_2$D $\rightarrow$ H + HDCS   &  HCO$^+$ + HDCS $\rightarrow$ H$_2$DCS$^+$ + CO     \\
&& HCO$^+$ + HDCS $\rightarrow$ H$_2$DCS$^+$ + CO &      &       p-H$_3$$^+$ + HDCS $\rightarrow$ H$_3$CS$^+$ + HD        \\
 \hline
\end{tabular}
%\end{center}
\label{mainreactions}
\end{table*}%
%\end{sidewaystable}

\begin{figure}
\centering
\includegraphics[width=0.5\textwidth]{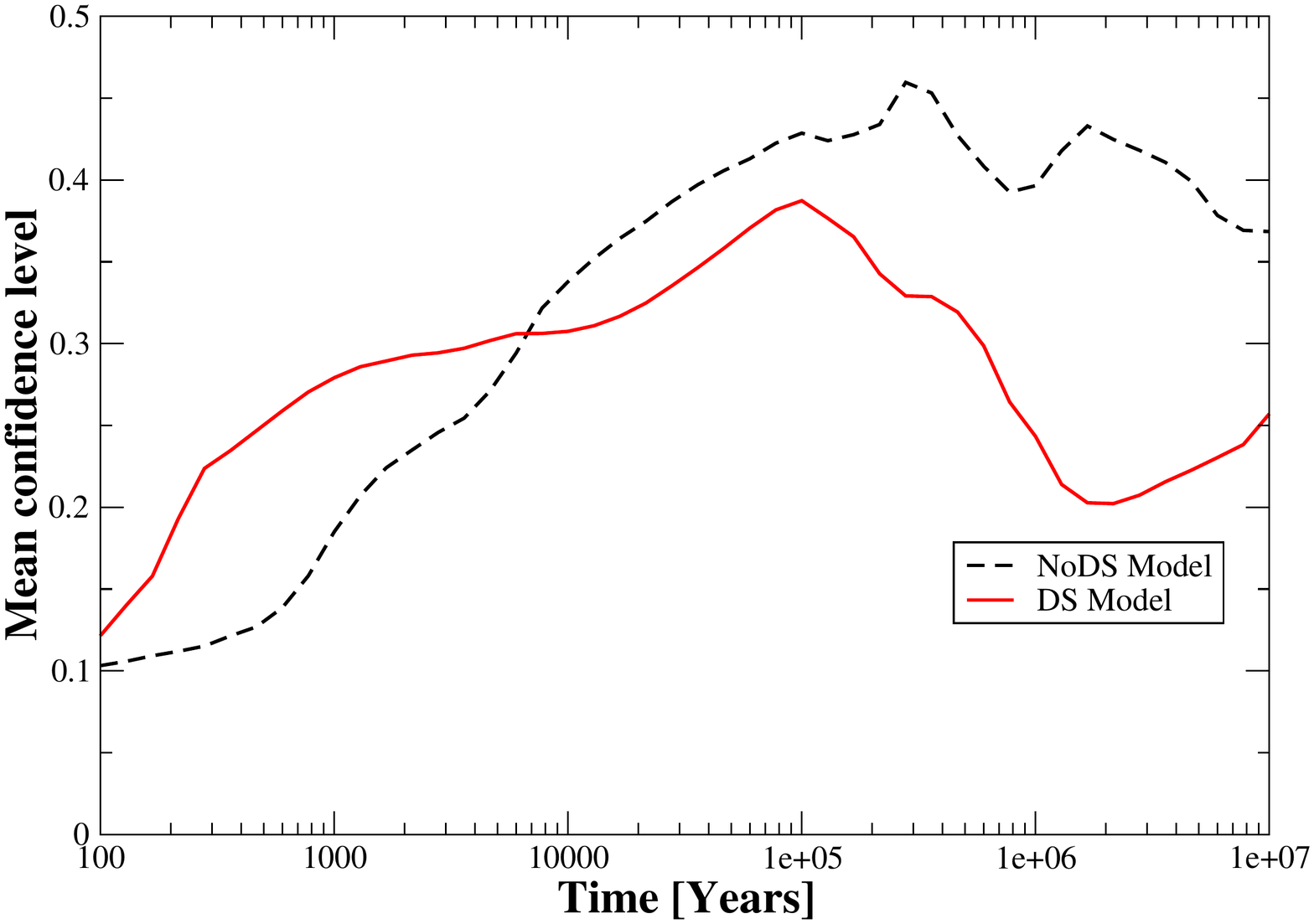}
\caption{Comparison between the modelled abundances for non deuterated species and the abundances observed in TMC-1 (see text). NoDS means the model without spin and deuterium chemistry whereas DS means the model with spin and deuterium chemistry.}
\label{fig:diffDist}
\end{figure}

\subsection{Gas phase D/H ratios in TMC-1: comparison with observations and previous models}

In the past, molecular D/H ratios observed towards TMC-1 were compared by various gas-grain chemical models (see for instance \citet{2000A&A...364..780R}, \citet{2013ApJS..207...27A}). 
\citet{2000A&A...364..780R} have studied the chemical evolution with deuterium fractionation for temperatures 10$-$100 K and densities $3\times 10^3-3\times 10^8$~cm$^{-3}$. They used a time-dependent gas-phase model based on the UMIST'95 database. Their chemical network consists of $\sim$ 300 species linked by $>$5\,000 reactions, and 
only includes singly deuterated species and limited surface chemistry for H$_2$ and HD. Recently, \citet{2013ApJS..207...27A} have also studied deuterium fractionation with their newly constructed chemical network, which contains 55000 reactions in gas phase. But they restricted their study to avoid reactions with -OH end groups and also they did not consider spin chemistry of H$_2$, H$_2$$^+$, and H$_3$$^+$ and their isotopologues. We did the same type of comparison but using our more complete model.
%For the first time, we have studied and compared observed molecular 
%D/H ratios in TMC-1 by not restricting the number of atoms in the deuterated species and by considering in detail the spin chemistry of  o-H$_2$, p-H$_2$, o-H$_2$$^+$, p-H$_2$$^+$, o-H$_3$$^+$, p-H$_3$$^+$ along with their isotopologues. We also consider most updated surface chemistry developed for studying ice deuteration by considering statistical branching ratios. 
In Table 3, we report molecular D/H ratios for several deuterated species observed in different positions of TMC-1 together with the abundances obtained with our model at $10^5$ yrs. For comparison, we also list the molecular D/H ratios reported by \citet{2000A&A...364..780R} and \citet{2013ApJS..207...27A} for the same physical conditions and the same age.

Our modelled abundances are in reasonable agreement with the observed ones. The D/H ratios predicted by our model are slightly smaller than the ones reported by \citet{2000A&A...364..780R}, \citet{2013ApJS..207...27A}. This can be explained by the fact that our network includes essential spin chemistry to take into account deuterium fractionation properly and the others not. Considering ortho and para species, reaction (1) becomes efficient in the backward direction. This results in efficient destruction of H$_2$D$^+$ 
which in turn globally reduces the abundances of deuterated species. This is the reason why we observe smaller molecular D/H ratios as compared to the other models.  

 \begin{table*}
\centering
\caption{Comparison of observed D/H ratios for different positions of TMC-1 cloud with predictions from our model (T = 10 K, $n_{\rm H} = 2\times10^{4}$ cm$^{-3}$, Time = $10^5$ year), \citet{2000A&A...361..388R} and  \citet{2013ApJS..207...27A} }. 
\begin{tabular}{c|ccccc}
\hline
\hline
%\tablewidth{0.45\textwidth}
Species		&	Our model	 &  Roberts \& Millar (2000a)          &   Albertsson et al. (2013)     & Observation & Observed Position \\
&    &   & & & \& Refs.\\
\hline
NH$_2$D/NH$_3$			        & 4.3 $\times 10^{-2}$	     &	8.4 $\times 10^{-2}$		      &	   5.2 $\times 10^{-2}$			& 	     9.0 $\times 10^{-3}$ - 1.4 $\times 10^{-2}$      &  TMC1-N (1)       \\
HDCO/H$_2$CO				& 2.1$\times 10^{-2}$		&	4.2 $\times 10^{-2}$		&	2.3 $\times 10^{-2}$		&         5.9 $\times 10^{-3}$ -  1.1 $\times 10^{-2}$    &      TMC1-CP (2 )\\
DCN/HCN					        & 0.4 $\times 10^{-2}$		&	0.9 $\times 10^{-2}$		&	2.4 $\times 10^{-2}$	&	   2.3 $\times 10^{-2}$  & 	TMC1-CP (3)\\
DNC/HNC					        & 	 0.6 $\times 10^{-2}$		&	1.5 $\times 10^{-2}$		&	1.6 $\times 10^{-2}$		&	 1.5 $\times 10^{-2}$  & TMC1-CP (2	)\\
C$_2$D/C$_2$H				& 0.4 $\times 10^{-2}$			&	1.1 $\times 10^{-2}$		&	1.5 $\times 10^{-2}$		&   1.0 $\times 10^{-2}$ &	TMC1-CP (4)	\\
C$_{4}$D/C$_4$H				&  2.3 $\times 10^{-3}$		&	0.4 $\times 10^{-2}$		&	1.0 $\times 10^{-2}$		&   4.0 $\times 10^{-3}$ &	TMC1-CP (5)	\\
DCO$^+$/HCO$^+$				& 0.3 $\times 10^{-2}$		&	1.9 $\times 10^{-2}$		&	1.8 $\times 10^{-2}$		&   	2.0 $\times 10^{-2}$ & TMC1-N (1)\\
N$_2$D$^+$/N$_2$H$^+$               &  0.2 $\times 10^{-2}$		 &	2.5 $\times 10^{-2}$		&	0.8 $\times 10^{-2}$		&		  8.0 $\times 10^{-2}$ & TMC1-N (1) \\
c-C$_3$HD/c-C$_3$H$_2$               & 1.6 $\times 10^{-2}$		&	0.6 $\times 10^{-2}$		&	1.3 $\times 10^{-2}$		& 	    8.0 $\times 10^{-2}$- 1.6 $\times 10^{-1}$ & TMC1-CP ( 6)    \\
C$_3$H$_3$D/C$_3$H$_4$		& 4.4 $\times 10^{-2}$	        &	   8.3 $\times 10^{-2}$		&	1.6 $\times 10^{-2}$		&	  5.4 $\times 10^{-2}$ -  6.5 $\times 10^{-2}$   &  TMC1-CP (7)    \\
DC$_3$N/HC$_3$N			        & 0.7 $\times 10^{-2}$		&	0.7 $\times 10^{-2}$		&	0.9 $\times 10^{-2}$			&   3.0 $\times 10^{-2}$ - 1.0 $\times 10^{-1}$ & TMC1-CP (8) \\
DC$_{5}$N/HC$_5$N			& 0.3 $\times 10^{-2}$		&	2.3 $\times 10^{-2}$		&	1.2 $\times 10^{-2}$		&  	1.3 $\times 10^{-2}$  &	TMC1-CP (9 )\\
HDCS/H$_2$CS				& 1.6 $\times 10^{-2}$	        &  	4.0 $\times 10^{-2}$		&	1.8 $\times 10^{-2}$		&		 2.0 $\times 10^{-2}$ &  TMC1-CP (10)   \\
c-C$_3$D$_2$/c-C$_3$HD                     & 0.5 $\times 10^{-2}$               &   -     & -&   4 $\times 10^{-2} $ & TMC1-C (11)  \\ 
\hline
\end{tabular} 
\footnotesize{
References: 1: \citet{2000A&A...356.1039T}; 2: \citet{2001ApJS..136..579T}; 3: \citet{1987IAUS..120..311W}; 4: \citet{1989ApJ...340..906M}; 5: \citet{1989ApJ...347L..39T}; 6: \citet{1988ApJ...326..924B}; 7: \citet{1992A&A...253L..29G}; 8: 
\citet{1994MNRAS.267...59H}; 9. \citet{1981ApJ...251L..33M}; 10. \citet{1997ApJ...491L..63M}; 12. \citet{2013ApJ...769L..19S}\\
Positions: TMC1-CP ($\alpha_{J2000}$ = 04$^h$ 41$^m$ 41$^s$.88, $\delta_{J2000}$ = +25$^{\circ}$ 41$^m$ 27$^s$), TMC1-N ($\alpha_{J2000}$ = 04$^h$ 41$^m$ 21$^s$.01, $\delta_{1950}$ = +25$^{\circ}$ 48$^m$ 11$^s$), 
TMC1-C ($\alpha_{J2000}$ = 04$^h$ 41$^m$ 16$^s$.1, $\delta_{J2000}$ = +25$^{\circ}$ 49$^m$ 43.8$^s$).\\
%Comments: According to \citet{2012A&A...546A.103S}, TMC1 has less advanced evolutionary in its south eastern part to more advanced in the north western part. 
}
\end{table*}

\begin{figure*}
\centering
\includegraphics[width=\textwidth]{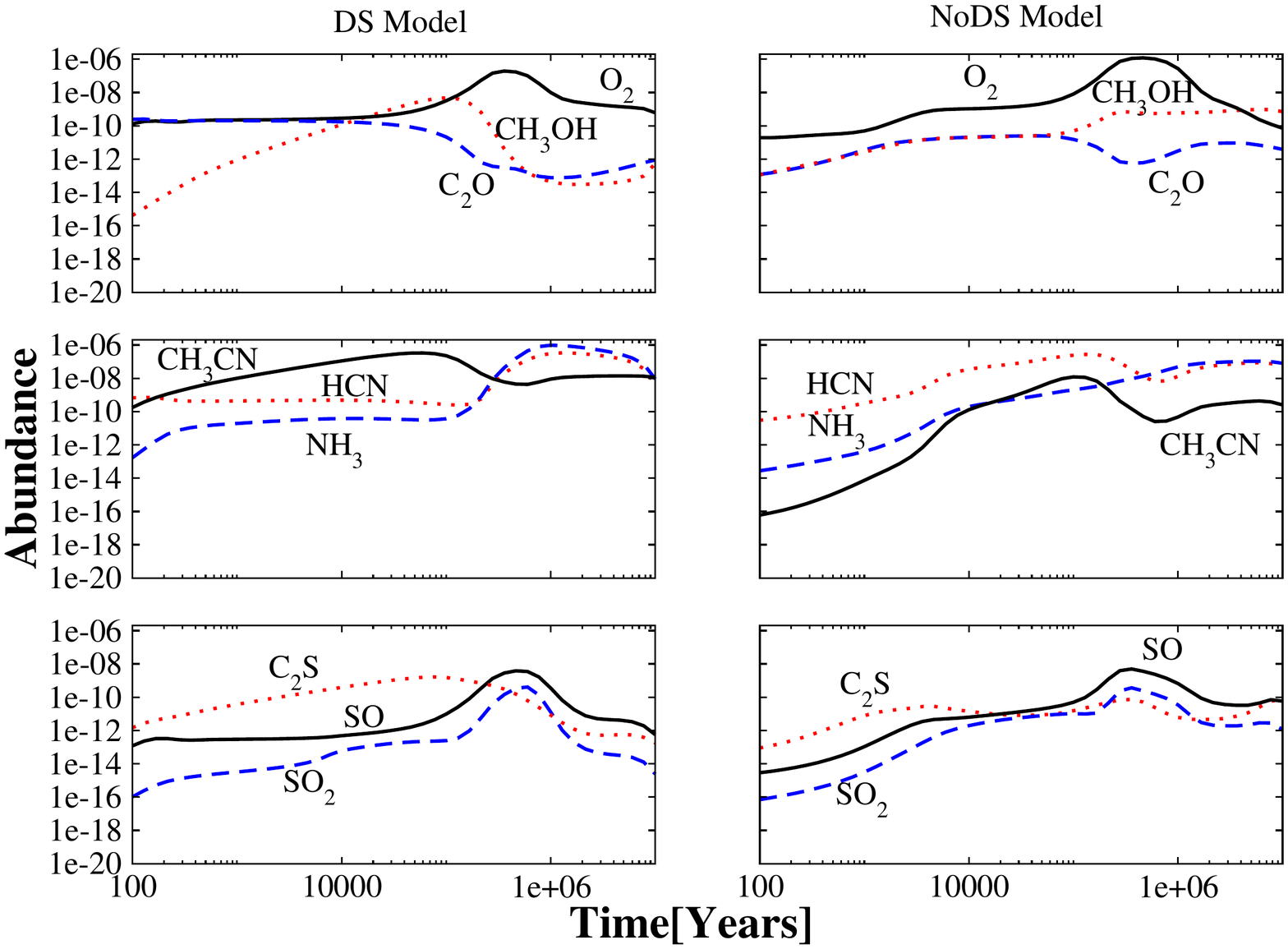}
\caption{Abundance with respect to n$_H$ for a selection of O, N, S bearing species and for two models: left panel: DS Model with spin and deuterium chemistry and right panel: NoDS model without spin and deuterium chemistry.}
\label{fig:diffDist}
\end{figure*}

\subsection{Effect of essential spin chemistry and deuteration on the observed non-deuterated species in TMC-1}

One of the goals of this study is to quantify the effect of essential spin chemistry and deuteration on the abundances of non-deuterated species. For that, we have run two models: (1) with deuteration and spin chemistry (DS) and (2) without deuteration and spin chemistry (NoDS). We then focus on the gas-phase species observed in the cold core TMC-1. In Table 4, we report the observed abundances in TMC-1 collected by \citet{2013ChRv..113.8710A} together with the abundances predicted 
by our two models. To define the ``chemical age" for TMC-1 for both models, we have calculated the mean confidence level following the prescription of \citet{2007A&A...467.1103G} to fit with the TMC-1 observations. 
First, we construct a log-normal distribution about each observational value, and then identify its defining standard deviation, $\sigma$, with an appropriate
error factor ({\em erf}) on the observed value. Finally,  the ``confidence", $\kappa_{i}$, that the modelled abundance ($X_{model,i}$, for species $i$) is {\em associated} with the observed value ($X_{obs,i}$) is defined as:
\begin{equation}
\kappa_{i} = \mbox{{\em erfc}} \left( \frac{| \log{\left(X_{model,i}\right)} - \log{\left(X_{obs,i}\right)} |}{\sqrt{2} \sigma} \right)
\end{equation}
\noindent where {\em erfc} is the complementary error function ({\em erfc} = 1 - {\em erf}); $\kappa_{i}$ ranges between zero and unity. In our case, we define $\sigma = 1$, hence 1 standard deviation corresponds to one order of magnitude higher or 
lower than the observed value. Fig. 5 represents the mean confidence level calculated using the method described above. As it can be seen, the DS model ($\sim$37\%) has a strong impact on the general confidence level as compared to the 
NoDS model ($\sim$45\%) at its maximum. The use of the DS model suggests a best fit chemical age for TMC-1 of 10$^5$ year whereas the NoDS model gives $3\times10^5$ year. At the time of best agreement, for both the DS and NoDS models, $\sim$50\% and $\sim$60\% of the observed species in TMC-1 are reproduced within factor of 10.  But all these estimates are very much dependent on the initial opr(H$_2$) and our estimates are only valid for the initial statistical ratio of 3:1. For example, in order to perform more realistic age estimates for TMC-1, it needs to be done in the context of parameter-space exploration with assumed opr(H$_2$) and this is out of scope of the current work. In Table 4, we have reported abundances for all the observed non-deuterated species using the above two models at their best chemical age. 

By investigating closely, we found three types of differences: (1) DS model results different from NoDS but in better agreement with the TMC-1 observation as compared to NoDS (Case I); (2) DS model results different from NoDS but in less agreement with the TMC-1 observation as compared to NoDS (Case II); (3) differences between the two models and the observations are both within a factor of 10 (Case III).The differences between the agreements are caused by changes in the chemical time scales. This can be clearly seen from Fig. 6 where we have shown the most affected oxygen, nitrogen and sulphur bearing species for which the chemical time scales for reaching their peak have changed. 
Here we will focus on the detailed chemistry for Case I and Case II, which shows the effect of deuteration and ortho-para chemistry on the observed non-deuterated species. For Case I, we found molecules such as O$_2$, CH$_3$OH, C$_2$O, H$_2$C$_4$, HC$_5$N, C$_2$S, C$_6$H$^-$, C$_8$H$^-$ and for Case II, we found CH$_3$C$_2$H, NH$_3$, HCN, CH$_3$CN, CH$_2$CHCN, SO, SO$_2$. The rest of the species belong to Case III.

Among the Case I species, O$_2$ forms via the gas phase reaction O + OH in both models. But in model DS, as can be seen from Table 4, the abundance of OH is almost 10 times lower than in the NoDS model, which in turn produces less O$_2$. 
In the DS  model, OH forms via the gas phase photodissociation of H$_2$O and by the neutral-neutral reaction O+ H$_2$CO$\rightarrow$H + OH + CO which has an activation barrier  while in the NoDS model, OH forms via the fast dissociative recombination of H$_3$O$^+$. CH$_3$OH forms via the dissociative recombination reaction of CH$_3$OH$_2$$^+$ in the gas phase at 10$^5$ years in the DS model whereas in the NoDS model, it forms via the 
surface reaction s-H + s-CH$_3$O and is destroyed rapidly via the barrier-less gas phase reaction CH$_3$OH + C. Reaction s-H + s-CH$_3$O becomes the main production reaction from 10$^6$ year in model DS i.e. well after the gas phase spin conversion of 
ortho to para H$_2$. C$_2$O is produced in model DS as well as in model NoDS via the gas phase reaction C + HCO. But we found that C$_2$O is more abundant in the DS model as compared to the NoDS model. This is due to the chemistry of HCO. HCO produced in  the NoDS model via the grain surface reaction s-H + s-CO which has a high activation barrier of 2500 K and in the DS model, it is produced via the barrier-less reaction CH$_2$ + O. We produce less C$_4$H$_2$ in model DS due to the highly efficient destruction of this species via the barrier-less gas phase reaction C + C$_4$H$_2$ whereas in the NoDS model, this destruction process is not very efficient. For HC$_5$N, we have better agreement with the observation and also it is more abundant in model DS than the NoDS. In both models, HC$_5$N forms via the dissociative recombination of H$_2$C$_5$N$^+$. But in model DS, HC$_5$N efficiently destroyed by p-H$_3$$^+$ to form H$_2$C$_5$N$^+$ which recombine again to form HC$_5$N. For C$_2$S, we observe that it is more abundant in DS model than NoDS model. This is due to the effect of the barrier-less gas phase reaction O + C$_2$S. This reaction destroys C$_2$S more rapidly in NoDS model as compared to the DS model. Among the anions, C$_6$H$^-$ and C$_8$H$^-$ both are efficiently produced in model DS as compared to NoDS. Since in model DS, the rate of production via the dissociative recombination reactions of  C$_6$H and C$_8$H  is very high due to high opr(H$_2$).  

Among the case II species, CH$_3$C$_2$H is produced efficiently in model NoDS due to its formation on the surface via the barrier-less reaction s-H + s-CH$_2$CCH.  Among the others, the chemistry of NH$_3$ and HCN is already been 
discussed in our earlier section. For CH$_3$CN, we observe a very high abundance in model DS as compared to the NoDS model. CH$_3$CN is produced by the dissociative recombination reaction of CH$_3$CNH$^+$ in both models but in model 
NoDS, it is destroyed by severals other ion-molecule reactions with HCO$^+$, H$_3$$^+$, H$_3$O$^+$, C$^+$ whereas in model DS, it is only destroyed by p-H$_3$$^+$. We observe that CH$_2$CHCN is less abundant in model DS  since it is 
destroyed by several reactions of ortho and para H$_3^+$ to form C$_3$H$_3$NH$^+$ whereas in the NoDS model, 
it is destroyed mainly by HCO$^+$. SO is mainly produced by the neutral-neutral reaction S + OH in model DS and OH is among the highly affected species in model DS. This corresponds to the decrease in abundance of SO in model DS. For SO$_2$, it is formed on the surface in DS model via the reaction s-O + s-SO whereas in model NoDS, it is formed by the neutral-neutral gas phase reactions OH + SO, O + SO. But its decrease in abundance in DS model is due to the efficient gas 
phase destruction reaction  C + SO$_2$.  The decrease in the abundance of NS in model DS as compared to NoDS is due to the efficient destruction by the various atoms i.e. C, O, N. 

\begin{table*}
\centering
\caption{Effect of deuteration and spin chemistry on molecular abundances (relative to H$_2$) at best chemical age obtained by comparing the models to the observed values in TMC-1}.
\begin{tabular}{c|ccccccc}
\hline
%\tablewidth{0.45\textwidth}
Species &	TMC-1	  &	NoDS	 & DS  & Species & TMC-1 &  NoDS & DS\\
\hline
 
 \hline
 OH                           & 3(-7)                                 &   3.2(-8)                     &   2.6(-9)                    &      *NH$_3$                     &  2.45(-8)            & 6.6(-9)                  &  3.7(-11)     \\
 
 H$_2$O                   & $<$7(-8)                           &     3.1(-7)                   &     1.8(-5)                  &     N$_2$H$^+$             &  2.8(-10)           &   7.9(-11)              &   7.5(-11)   \\

 *O$_2$                     & $<$7.7(-8)                         &  7.3(-7)                     &       3.2(-9)                &     CN                            & 7.4(-10)              &    2.7(-8)               &       9.4(-9)  \\

 CO                           & 1.7(-4)                              &     6.0(-5)                  &      6.7(-5)                 &      *HCN                       &  1.1(-8)                &     8.7(-8)             &  2.6(-10) \\

HCO$^+$                 &  9.3(-9)                              &       4.2(-9)                &       1.0(-9)               &       HNC                        &  2.6(-8)                 &        3.1(-8)          &       4.8(-8)\\

H$_2$CO                &  5(-8)                                  &         1.2(-8)             &    1.4(-9)                  &                                        &                              &                            &                \\

*CH$_3$OH              & 3.2(-9)                                &      7.3(-10)              &      4.8(-9)               &                                        &                               &                           &                \\
 
                                 &                                           &                                  &                              &   *CH$_3$CN                     &  6(-10)                     &     4.2(-10)         &    2.2(-7)  \\

*C$_2$O                 & 6(-11)                                   &        6.9(-13)               &     2.1(-11)             &  CH$_2$CHCN             &1(-9)                           &   1.2(-11)         & 1.3(-12)       \\

CH$_2$CO           &  6(-10)                                    &          9.0(-10)            &    2.0(-10)              &  C$_3$N                    & 6(-10)                            &  1.9(-9)           & 8.6(-10)       \\

CH$_3$CHO        & 6(-10)                                    &        2.3(-12)                &    5.3(-12)              &  HC$_3$N                 &  1.6(-8)                           &   1.0(-8)        &  1.0(-7)    \\

C$_3$O                & 1(-10)                                &       7.7(-9)                   &     7.9(-10)                     &      HNC$_3$            &   3.8(-11)                         &  4.1(-10)        & 3.6(-10) \\

CH                       &  2.0(-8)                                &     2.0(-9)                  &       3.5(-9)                      &                                 &                                         &                       &  \\

C$_2$H               &  7.2(-9)                               &         3.1(-9)              &       1.5(-9)                       &                                &                                         &                         &      \\

$c$-C$_3$H        &  1.03(-9)                        &         3.2(-9)                 &       2.8(-9)           &        & &&   \\

$l$-C$_3$H          &  8.4(-11)                    &           2.4(-9)               &          1.5(-9)         & C$_5$N                   &  3.1(-11)        &  2.1(-10)  & 5.6(-11) \\

$c$-C$_3$H$_2$  & 5.8(-9)                        &             2.6(-8)            &           2.6(-10)         &  HC$_5$N               & 4(-9)      &   4.8(-10) &    1.03(-9)  \\

$l$-C$_3$H$_2$   &  2.1(-10)                      &             2.1(-8)              &             9.2(-9)       &  CH$_3$C$_5$N    &  7.4(-11)        &   4.8(-13)        &  5.1(-12) \\

CH$_3$C$_2$H      &  6(-9)                            &         2.7(-10)             &          8.4(-12)          &  HC$_7$N              & 1(-9)          &     3.9(-11)           & 1.7(-10)    \\

                                  &          &                 &                      &  HC$_9$N              &  5(-10)         & 1.3(-12)  & 5.8(-11)      \\

C$_4$H                  &  7.1(-8)                         &      1.0(-9)        &              8.0(-10)           &                & &      &    \\

C$_4$H$^-$          &  $<$3.7(-12)            &         1.7(-11)     &           1.9(-11)              &  NO                        &  2.7(-8)         & 7.0(-8)   &   6.8(-10)   \\

H$_2$C$_4$          &  7(-10)                           &            2.7(-8)  &               1.9(-11)           & HNCO                   & 4(-10)               &  3.2(-13) &  4.0(-11)\\

C$_5$H                  & 8(-10)                    &          5.2(-10)    &              1.0(-10)            &               &  &   &   \\
 
CH$_3$C$_4$H     &  1(-9)                      &           6.1(-11)    &           3.8(-11)              &                              &   &&  \\

C$_6$H                  &  4.1(-10)                   &     3.7(-11)             &         3.7(-11)             &  H$_2$S                 &  $<$5(-10)    &       4.5(-11)   &    2.1(-10) \\

C$_6$H$^-$          &  1.0(-11)               &       4.2(-12)          &            1.0(-11)           &  CS                         &  2.9(-9)           &  2.6(-9)  & 2.2(-9)    \\

H$_2$C$_6$          &  4.7(-11)       &  9.7(-11)         &           6.8(-12)           &  HCS$^+$              &  3(-10)            &   2.7(-12)    &   2.0(-12)   \\

C$_8$H                  & 4.6(-11)           &    9.3(12)      &             1.4(-11)         & H$_2$CS               &  7(-10)          & 1.37(-10)    &   2.1(-10)   \\

C$_8$H$^-$          & 2.1(-12)         &      8.1(-13)   &           2.3(-12)          &  *C$_2$S                &  7(-9)            & 6.8(-11)  &   1.5(-9) \\

  &       &         &                    & C$_3$S                  & 1(-9)               &  6.2(-10)  & 4.7(-10) \\

OCS                       &  2.2(-9)                                  &   2.7(-10)      &             6.0(-10)          &   *SO                        &  1.5(-9)        & 3.9(-9)  & 9.6(-12)    \\
 
NS                         &  8.0(-10)                             &      3.6(-11)   &          1.6(-13)            & *SO$_2$                  & 3(-10)         & 2.4(-10)  & 2.4(-13)   \\
\hline
\end{tabular}
\footnotesize{
\,\\
a(b) refers to a $\times$ 10$^{b}$. \\
Abundances correspond to the positions TMC-1 $\alpha_{J2000}$ = 04$^h$ 41$^m$ 41$^s$.88, $\delta_{J2000}$ = +25$^{\circ}$ 41$^m$ 27$^s$ (cyanopolyyne peak). See Agundez \& Wakelam (2013) for more details.\\
DS refers to modelled abundances with deuteration and spin chemistry taken into account\\
NoDS refers to modelled abundances without deuteration and spin chemistry\\
* Fig. 6 shows that the shift in the chemical time scale for the affected species (bold face) to reach the peak abundances.}
\end{table*}

\subsection{Some new observable deuterated species in TMC-1 from our model}
In Table 5, we list a set of potentially new observable deuterated species in TMC-1. For predicting the observability (in the frequency range of 73- 180 GHz) of all 
the species (for which frequencies are known) listed in Table 5, we have used the CASSIS\footnote{\url{http://cassis.irap.omp.eu}} interactive spectrum analyser to calculate the line intensities under local thermodynamic 
equilibrium (LTE) conditions (though LTE might not be reasonable for some of those transitions at 10$^4$ cm$^{-3}$, see for other alternatives in \citet{2016ApJ...823..124L}). For our calculations, we assume excitation 
temperature= 5 K (to take into account the fact that the populations of the molecular levels might not be at LTE), FWHM= 1 km/s, H$_2$ column density = 10$^{22}$ cm$^{-2}$ and no beam dilution to simulate typical 
dark cloud conditions. We list only the brightest transitions along with their Einstein coefficients and upper level energies in Table 5. 

\begin{table*}
\centering
\caption{New set of observable gas phase deuterated species in TMC-1 awaiting detection (for example with the IRAM-30m telescope).}
\begin{tabular}{lcccccc}	
\hline
\hline
%Species			&	 n(x)/n(H) 	& 	Line		& 	Frequency [GHz]	& 	Line Intensity [K]	& A$_{ij}$ [s$^{-1}$] & E$_{up}$ [K]	\\
Species & Line	& A$_{ij}$ [s$^{-1}$] &  E$_{up}$ [K] &  Frequency [GHz]	&  n(x)/n(H) & Line Intensity [K] \\
\hline
%c-C$_3$D		               &	2.0(-10)	&	3 1 3 	$\rightarrow$~2 1 2		&	116.56 (JPL)			&	0.5		&	2.2(-5) &	9.39 \\

c-C$_3$D	 & 3 1 3 	$\rightarrow$~2 1 2 &	2.2(-7) &	9.41 & 116.73 (JPL)	&2.0(-10)	& 0.18\\
\hline
%l-C$_3$HD			&		1.4(-10)						&	5 0 4 	$\rightarrow$~4 0 4	      	&		96.90	 (CDMS)	&	 	0.17	&	8.0(-5)  &	 13.95\\
l-C$_3$HD			&	4 0 4 	$\rightarrow$~3 0 3 &	4.05(-5)  &	 9.30 & 	77.52	 (CDMS) & 1.4(-10)	&	0.055	\\
\hline

%CH$_2$ND	       &	 3.2(-11)							&	3 0 3 	$\rightarrow$~2 0 2			&		175.18	(JPL) 	&	0.01	&	7.3(-6) &  16.84 \\

CH$_2$ND & 2 0 2 	$\rightarrow$~1 0 1 & 5.4(-6) &  8.43 & 117.08	(JPL) &  3.2(-11)		&  0.005\\
\hline
%DNCCC	       &	5.2(-11)							&	 9 9 9 	$\rightarrow$~8 7 8		&		79.20 (JPL)	&		0.16	&  1.9(-8)	&   19.01\\

DNCCC	       &  9 8 7 	$\rightarrow$~8 8 8 & 1.9(-8)	&   19.01 & 79.20 (JPL) &  5.2(-11) &  0.019\\

\hline
%C$_4$HD				&	2.3(-9)				&		9 0 9$\rightarrow$~ 8 0 8 				&	76.27 (CDMS)	&	0.98  & 4.11(-5)	 &   18.30\\

C$_4$HD	& 9 0 9$\rightarrow$~ 8 0 8 	&  4.11(-5)	 &   18.30 & 76.27 (CDMS) & 2.3(-9)	& 0.13\\

%\hline
%C$_5$D			       &	1.9(-11)						&		16 1 16  $\rightarrow$~ 15 1 15         	&		74.41 (JPL)		&	0.005 			&   5.5(-5)	& 31.16	\\

%C$_5$D	& 16 1 16  $\rightarrow$~ 15 1 15 & 5.5(-5)	& 31.16 &  74.41 (JPL) & 1.9(-11)	& 0.005\\
\hline
%DNO & 1.7(-11) &		2 0 2   $\rightarrow$~ 1 0 1         	&		146.116 (JPL)		&	0..01 			&   8.4(-6)	&   10.52 \\
DNO & 2 0 2   $\rightarrow$~ 1 0 1 & 5.0(-6)	&   10.52 & 146.117 (JPL) & 1.7(-11) & 0.003 \\
\hline
\hline
\multicolumn{7}{l}{The table is limited to species with relative abundances $\geq 10^{-11}$.}\\
\multicolumn{7}{l} {a(b) refers to a $\times$ 10$^{b}$. }\\ 
\end{tabular}
\end{table*}

\section{Conclusions}
In this work, we present the first publicly available chemical network of multiply deuterated species along with spin chemistry of H$_2$, H$_2$$^+$, H$_3$$^+$ and their isotopologues. 
We also benchmarked our new model by comparing with existing works. Observed 
molecular D/H ratios at different positions of TMC-1 have been compared with our new comprehensive astrochemical model. By introducing an extensive description of deuteration along with nuclear spin state processes, we found 
that the chemical time scale for reaching the peak abundance of some non-deuterated species has changed by a factor of a few in some cases. Finally, we report a new set of potentially observable deuterated species in TMC-1. 

\section*{Acknowledgements}

LM, PG, MR, VW, FH thanks ERC starting grant (3DICE, grant agreement 336474) for funding during this 
work. PG postdoctoral position is funded by the INSU/CNRS. VW, AD, SG, FH also acknowledge the CNRS programme PCMI for funding of their research. We would like to thank the anonymous referee for constructive comments 
that helped to improve the manuscript.

\bibliographystyle{mnras}

\bibliography{Deuteration}

\end{document}